\documentclass[
 reprint,
bibnotes,
 amsmath,amssymb,
 aps,
pra,
]{revtex4-2}
\usepackage{natbib}
\usepackage{graphicx} 
\usepackage{amsmath}
\usepackage{amssymb}
\usepackage{caption}
\usepackage{subcaption}
\usepackage{amssymb}
\usepackage{xcolor}
\usepackage{comment}
\usepackage{float}

\begin{document}
\title{Particle transport in a correlated ratchet}

\author{Saloni Saxena}
\affiliation{Max Planck Institute for the Physics of Complex Systems, Dresden, Germany}
\author{Marko Popovi\'c}
\affiliation{Max Planck Institute for the Physics of Complex Systems, Dresden, Germany}

\author{Frank J\"ulicher}
\affiliation{Max Planck Institute for the Physics of Complex Systems, Dresden, Germany}

\begin{abstract}
One of the many measures of the non-equilibrium nature of a system is the existence of a non-zero steady state current which is especially relevant for many biological systems. To this end, we study the non-equilibrium dynamics of a particle moving in a tilted colored noise ratchet in two different situations. In the first, the colored noise variable is reset to a specific value every time the particle transitions from one well to another in the ratchet. Contrary to intuition, we find that the current magnitude decreases as the correlation time of the noise increases, and increases monotonically with noise strength. The average displacement of the particle is against the tilt, which implies that the particle performs work. We then consider a variation of the same problem in which the colored noise process is allowed to evolve freely without any resetting at the transitions. Again, the average displacement is against the potential. However, the current magnitude increases with the correlation time, and there is an optimal noise strength that maximizes the current magnitude. Finally, we provide quantitative arguments to explain these findings and their relevance to active biological matter such as tissues.

\end{abstract}

\maketitle

\section{Introduction} 

Active matter consists of components that use energy from the environment to produce self-propelled motion \cite{marchetti}. Biological systems such as swarms of bacteria, flocks of birds and tissues are well-known examples. Self-propulsion ensures that the system operates far from equilibrium. 

Several tools have been used to quantify non-equilibrium systems, such as violation of the fluctuation-dissipation theorem, entropy production rate and steady-state current flows \cite{tchou}. A specific case which has attracted much interest in the last few decades is that of a ratchet. A ratchet is a system consisting of a particle moving in a periodic potential with identical asymmetric wells \cite{reimann}. The particle is subjected to noise which may be white or colored. The asymmetric nature of the wells causes the particle to move in a preferred direction and results in a non-zero current.

\begin{figure}[H]
\includegraphics[width=0.45\textwidth]{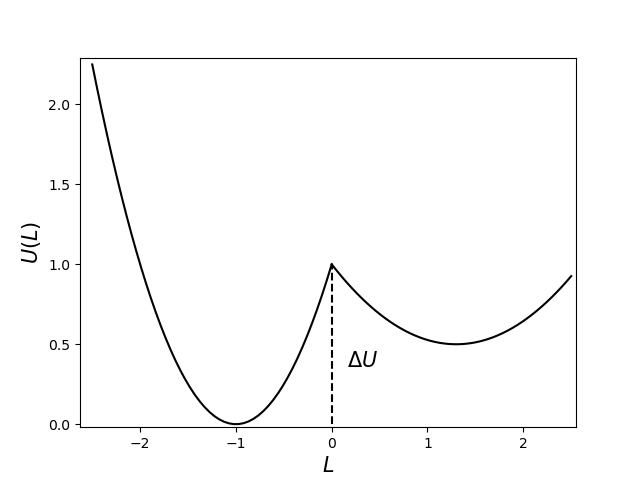}
\caption{Schematic representation of the potential landscape in tissues. $L$ is the bond length, which becomes zero at the T1 transition. Notice the cusp at $L=0$.}
\label{fig: cusp}
\end{figure}

In this work, we use a ratchet model to investigate steady state currents and work in active matter. This question is particularly relevant in biological tissues. Tissue dynamics are governed by processes such as chemical signalling between cells and force generation by collective movement \cite{wang, bosveld}. Force generation can occur through a variety of mechanisms such as cell elongation, cell division, cell extrusion and T1 transitions \cite{duclut}. Of particular interest are T1 transitions, in which the bond between two neighboring cells shrinks until it vanishes and a new bond is formed in the transverse direction \cite{alt}. They are well-known in the study of passive systems such as foams where they dissipate external stresses \cite{cohenaddad, biance}. In a tissue, however, T1 transitions occur along an axis set by anisotropic processes occurring within it, resulting in contractility in a preferred direction \cite{duclut, keller, tada}. This can lead to accumulation of stresses rather than dissipation, implying that work is done by T1 transitions. 

This paper is loosely motivated by understanding how T1 transitions do work; it should be noted, however, that our approach can be applied to any non-equilibrium system regardless of T1 transitions.
We construct a toy model consisting of a particle moving in a ratchet. To ensure that the system is out-of-equilibrium, we subject the particle to colored noise. We choose the shape of the ratchet potential based on the properties of common models of tissues such as the vertex model \cite{alt}, described in detail below. 

\begin{figure*}
     \centering
     \begin{subfigure}[b]{0.4\textwidth}
         \centering
         \includegraphics[width=\textwidth]{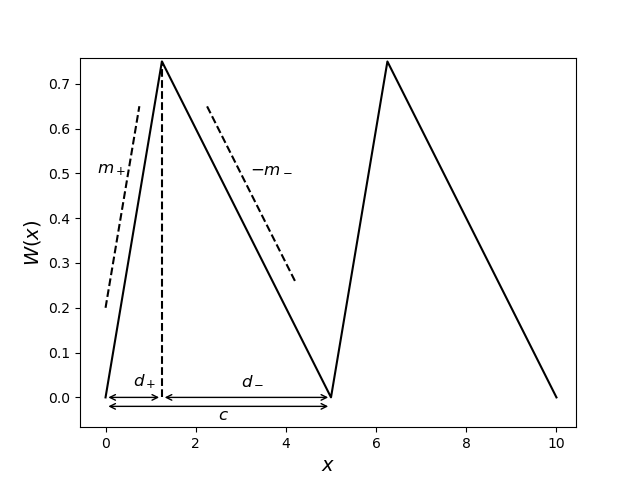}
         \end{subfigure}
     \begin{subfigure}[b]{0.4\textwidth}
         \centering
         \includegraphics[width=\textwidth]{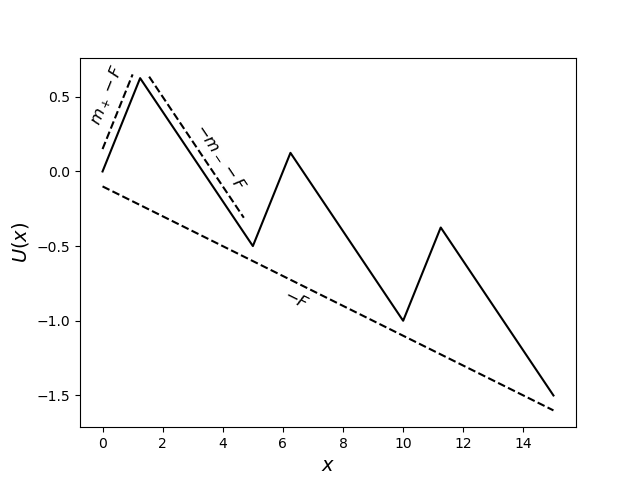}
         \end{subfigure}
         
    \caption{Schematic of the model potential. (a) A flat sawtooth potential, with slope $m_+$ on the short side and slope $-m_-$ on the long side. The projections of the short and long sides on the X axis are $d_+$ and $d_-$. The base of the triangles forming the sawtooth is $d_+ + d_-=c$. (b) After tilting, the slopes of the short and long sides are $m_+-F$ and $m_--F$.}
    \label{fig: potential}
    \end{figure*}

The rest of this paper is organized as follows: \\
In Section \ref{section: corr}, we briefly describe the vertex model of tissues, specifically its energy landscape and how it incorporates T1 transitions. We then introduce our model and the associated stochastic differntial equations. We show our results in Sections \ref{section: resetting} and \ref{section: no-reset}. In Sec. \ref{section: resetting}, we look at a ratchet where the colored noise process is reset to some value every time the particle jumps from one well to another. In Sec. \ref{section: no-reset}, we use the same ratchet, but allow the colored noise to evolve freely without resetting. In both cases, the current is calculated for various model parameters and comparisons are made between the two cases. Finally, we summarize our findings and suggest avenues for future work in Section \ref{sec: summary}.

\section{One dimensional correlation ratchet} \label{section: corr}

\subsection{Models of Tissue Dynamics}
Epithelial tissues can be represented as a tight network of polygons which are joined to one another by common edges meeting at vertices. The dynamical properties of such a system are
described by what is known as the vertex model \cite{bi3, alt, fletch, tetley, commelles, yama}.
It is a many-particle model containing interactions between large numbers of vertices via a set of bonds. The energy landscape of this model is made up of contributions due to cell area changes, cell perimeter changes and bond tensions. The bond tensions are time-dependent stochastic variables and are usually represented by an Ornstein-Uhlenbeck (OU) process. The OU process is temporally correlated i.e. it represents colored noise. A schematic of the energy landscape is shown in Fig. \ref{fig: cusp}. Most importantly, the energy landscape shows a cusp when a T1 transition occurs, which is also shown in Fig. \ref{fig: cusp}

Based on these properties, we choose a correlation ratchet consisting of a sawtooth potential, as shown in Fig. \ref{fig: potential}(a).  
Each well in the potential consists of two sides with different slopes. Further asymmetry is introduced by imposing an overall tilt obtained by subtracting a linear term from the flat potential in Fig. \ref{fig: potential}(a). This yields the tilted correlation ratchet shown in Fig. \ref{fig: potential}(b). Note that this potential already incorporates cusps.

The relevant parameters of the potential are as follows: The flat potential of Fig. \ref{fig: potential}(a), which we call $W(x)$ has slopes $m_+$ for the short side and $-m_-$ for the long side. $m_+$ and $m_-$ are positive, with $m_+ > m_-$. $d_+$ and $d_-$ are the horizontal distances traveled when the particle moves along the short and long sides respectively. The sum of $d_+$ and $d_-$ is the length of the base of each triangle and is given by $c$. We subtract a term $Fx$ from $W(x)$ to obtain the final potential of Fig. \ref{fig: potential}(b). The slopes of the sides then become $m_+-F$ and $-m_- - F$ and the resulting potential is denoted by $U(x)$. 

\begin{equation} \label{eq: tilt}
    U(x) = W(x)-Fx
\end{equation}

The motion of a particle in this potential, subjected to colored noise is given by the following Langevin equation
\begin{equation} \label{eq: lang_x}
    \dot{x} = -U^\prime(x) + \lambda(t)
\end{equation}

\begin{equation}
    U^\prime(x) = 
\begin{cases}
m_+ - F, \ \text{short side} \\
-m_- -F, \ \text{long side} \\
\end{cases}
\end{equation}

$\lambda(t)$ is the OU process, which evolves according to

\begin{align} \label{eq: ou}
    \dot \lambda(t) &= -\frac{\lambda(t)}{\tau} + \eta(t) \\
    \langle \eta(t) \rangle &= 0 \\
    \langle \eta(t) \eta(t^\prime) \rangle &= \frac{\sigma^2}{\tau} \delta(t-t^\prime)
\end{align}
$\tau$ is the correlation time of $\lambda$ which sets the time scale over which two-time correlations in $\lambda$ decay. 
\begin{equation}
    \langle \lambda(t) \lambda(t^\prime) \rangle = \frac{\sigma^2}{2} e^{-(t-t^\prime)/\tau}
\end{equation}
The probability density function for the OU process with initial value $\lambda(t=0) = \lambda_0$ is \cite{gardiner}

\begin{equation} \label{eq: prob_lbda}
    P(\lambda, t) = \frac{1}{\sqrt{\pi \sigma^2 (1-e^{-2t/\tau})}} \ \text{exp} \left[ -\frac{\lambda^2}{\sigma^2 (1-e^{-2t/\tau})} \right]
\end{equation}

We can define a ``well'' as any region bounded by a short side on the right and a long side on the left. The wells are denoted by $w_i(t)$ from left to right, with the index $i$ being an integer. In this notation, a downward jump to the next well corresponds to an increase in $x$ (and hence in $w$) and an upward jump to a decrease in $x$ and $w$
.
Unless mentioned otherwise, we take $m_+ = 2$, $m_- = 0.05$, and $F=0.1$ for our simulations. The base of each triangle is $c=d_+ + d_-=5$. We choose $d_-$ to be the unit of distance in what follows. It sets the length scale that the particle must travel to move to the previous well.
We also define a characteristic time scale $t_c$ associated with relaxation in a single potential well. We choose $t_c$ to be the time taken for the particle to slide down the long side in the absence of noise. Hence, 
\begin{equation}
   t_c = \frac{d_-}{m_- + F} 
\end{equation}
In what follows, all times are defined in units of $t_c$.

\section{Resetting noise at each jump} \label{section: resetting}

In order to make this toy model a better representation of tissues, we implement the following protocol: Equations (\ref{eq: lang_x}) and (\ref{eq: ou}) are integrated using the Euler-Maruyama method \cite{kloeden}. The well number corresponding to the instantaneous position of the particle is recorded at each time step. Whenever the particle jumps to a different well, we reset $\lambda(t)$ to zero, while $x(t)$ continues to evolve as usual. Since the probability distribution function of $\lambda$ is symmetric about zero, resetting to $\lambda=0$ ensures that no additional asymmetry is introduced via the resetting process. 

First, it is useful to draw some qualitative conclusions about the dynamics from Eq. (\ref{eq: lang_x}). For the particle to make an upward jump, the particle has to travel along the long side and over the cusp. Since $x(t)$ decreases in this process, we require that $\dot x(t)$ be negative. From the equation for the long side, we have

\begin{equation}
    \lambda < -m_- - F = \lambda_b
\end{equation}

\begin{figure*}
     \centering
     \begin{subfigure}[b]{0.45\textwidth}
         \centering
         \includegraphics[width=\textwidth]{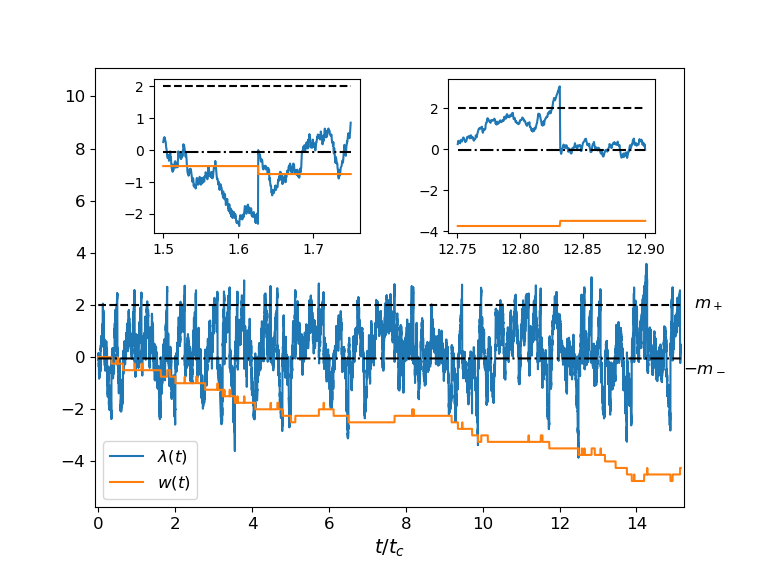}
         \end{subfigure}
     \begin{subfigure}[b]{0.45\textwidth}
         \centering
         \includegraphics[width=\textwidth]{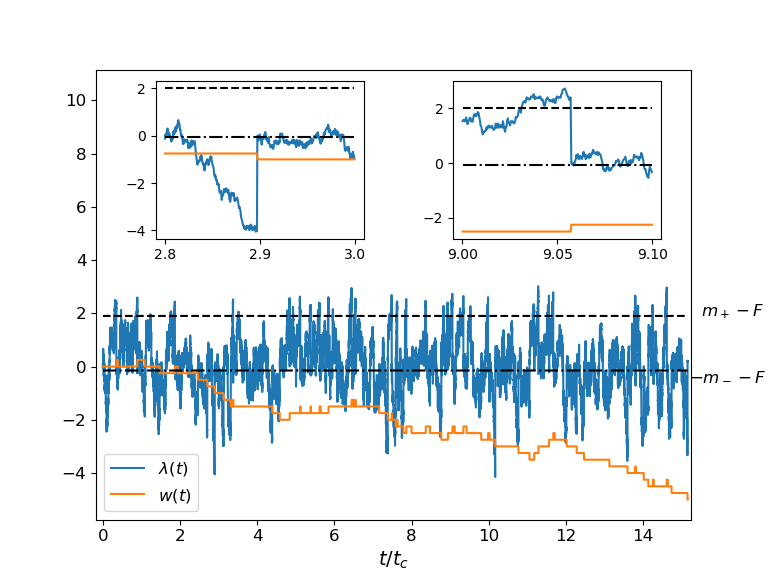}
         \end{subfigure}
     
    \caption{Snapshots showing a typical trajectory of $w(t)$ and $\lambda(t)$, (a) without tilting (b) with tilting. The blue curve is $\lambda(t)$ and the orange curve is $w(t)$ which is the instantaneous well number. In both (a) and (b), the inset on the left is an example time interval $T_-$ where $\lambda(t)<\lambda_b$ (dash-dotted line). Here, $w(t)$ decreases with time, corresponding to upward motion. The right inset is an example time interval $T_+$ where $\lambda(t)>\lambda_f$ (dashed line). Well number $w(t)$ increases with time, corresponding to downward motion.}
    \label{fig: one_traj_res}
    \end{figure*}

Similarly, for a downward jump we have
\begin{equation}
    \lambda > m_+ - F = \lambda_f
\end{equation}
$\lambda_b \ (\lambda_f)$ give an upper (lower) bound on the value of $\lambda$ needed to jump up (down). Note that if the potential were not tilted, these thresholds would simply be $-m_-$ and $m_+$. For $m_+ = 2$, $m_-=0.05$ and $F=0.1$, $\lambda_b=-0.15$ and $\lambda_f=1.9$
We show a single trajectory in Fig. \ref{fig: one_traj_res}. Fig. \ref{fig: one_traj_res}(a) has no tilt ($F=0$) and \ref{fig: one_traj_res} (b) has a non-zero tilt with $F=0.1$. 

The blue curve represents $\lambda(t)$ and the orange curve represents the well number $w(t)$. The well number has been rescaled to be of the same order of magnitude as $\lambda$ so that both can be seen clearly on the same plot. In both figures, the dashed line represents $\lambda_b$ and the dash-dotted line represents $\lambda_f$. The insets show intervals where $\lambda$ is less than $\lambda_b$ (right) and $\lambda > \lambda_f$ (left). These time intervals are denoted by $T_-$ and $T_+$ respectively. In both the flat and tilted cases, we can see that jumps occur shortly after these threshold values of $\lambda$ are crossed. 

\begin{figure} 
     \centering
     \begin{subfigure}[b]{0.45\textwidth}
         \centering
         \includegraphics[width=\textwidth]{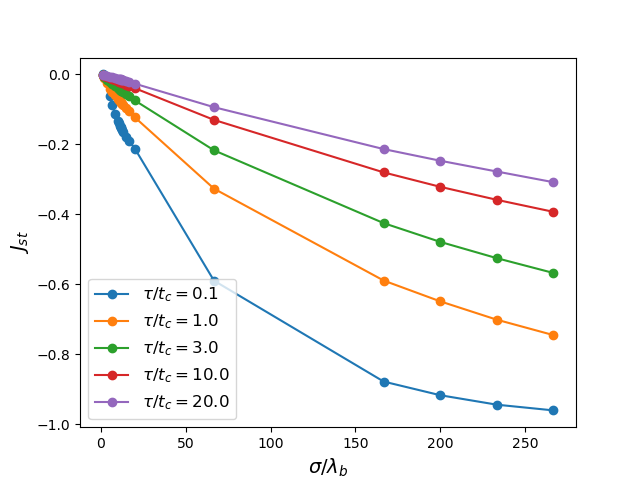}
         \end{subfigure}
    
    \caption{$J_{st}$ as a function of $\sigma$ when $\lambda$ is reset to zero after every jump. It \textit{decreases} with $\tau$ and increases with noise strength for fixed $\tau$. } 
    \label{fig: current_reset}
    \end{figure}

We now compute the steady-state particle current $J_{st}$. It is defined as \cite{reimann}
\begin{equation} \label{eq: current}
    J_{st}(t) = \lim_{t \rightarrow \infty} \langle \dot x(t) \rangle
\end{equation}
where the angular brackets $\langle \ \rangle$ denote an ensemble average over trajectories. The sign of $J_{st}$ indicates the direction of the net particle motion. $J_{st}$ is shown as a function of the applied noise in Fig. \ref{fig: current_reset} for different values of the correlation time $\tau$. 

\begin{figure*}
     \centering
     \begin{subfigure}[b]{0.3\textwidth}
         \centering
         \includegraphics[width=\textwidth]{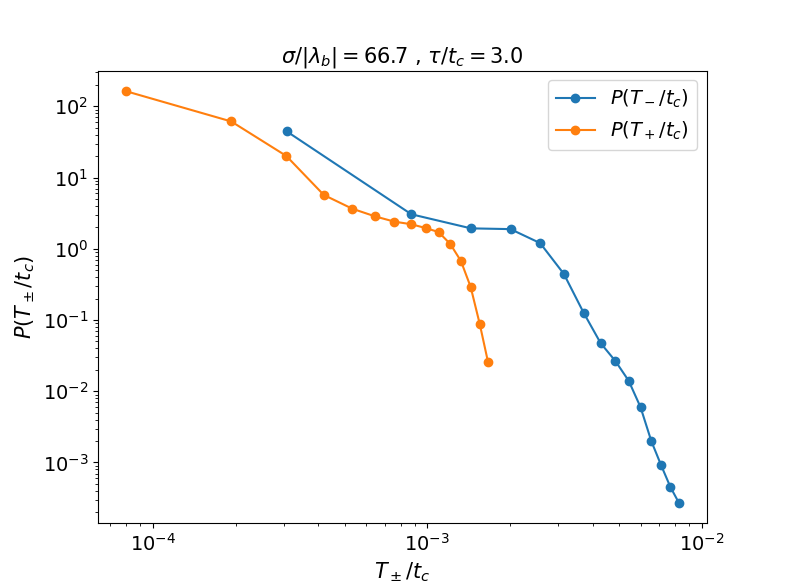}
         \end{subfigure}
     \begin{subfigure}[b]{0.3\textwidth}
         \centering
         \includegraphics[width=\textwidth]{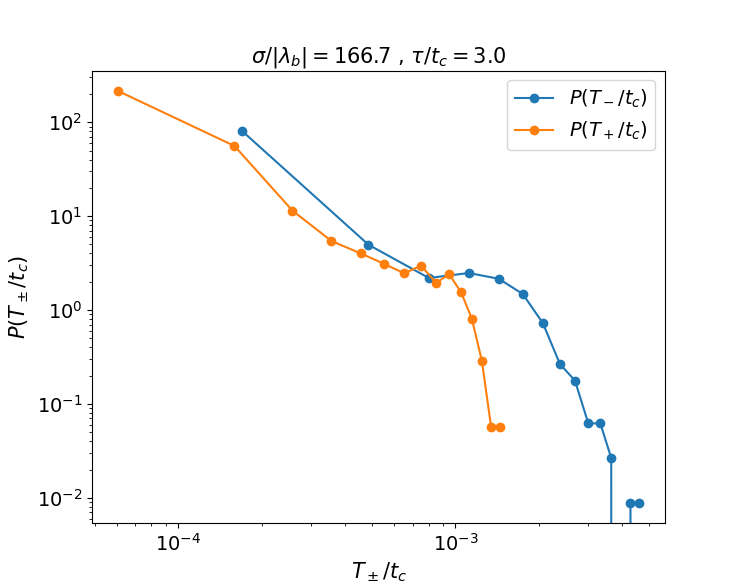}
         \end{subfigure}
    \begin{subfigure}[b]{0.3\textwidth}
         \centering
         \includegraphics[width=\textwidth]{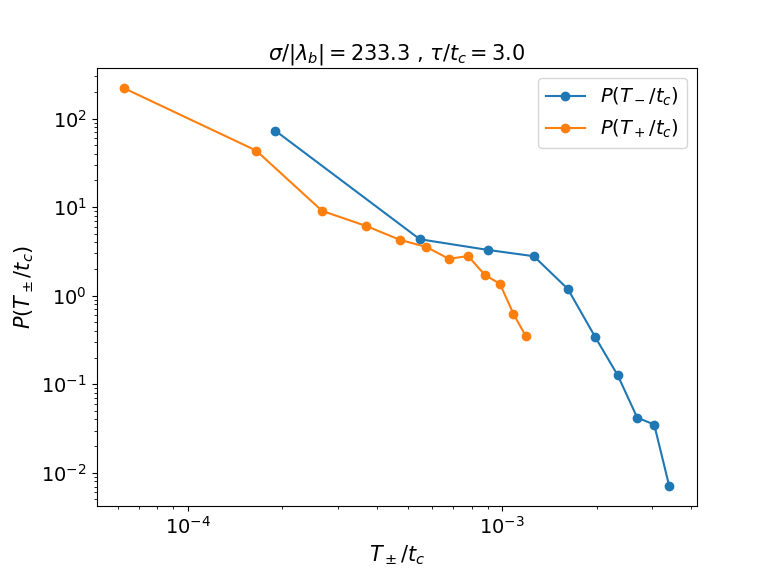}
         \end{subfigure}     
    \caption{Distribution of $P(T_\pm/t_c)$ for $\tau=3 \ t_c$ and (a) $\sigma=66.7 \ |\lambda_b|$, (b) $\sigma=166.7 \ |\lambda_b|$, (c) $\sigma=233.3 \ |\lambda_b|$. Even for the largest noise, large values of $T_-$ are more likely than large values of $T_+$ and hence the average number of upward jumps is greater than that of downward jumps.}
    \label{fig: res_fpt_dist}
    \end{figure*}
    
The steady state current in each case is negative, indicating that the average particle velocity is negative and that the particle moves up the potential on average. The current magnitude $|J_{st}|$ decreases with $\tau$.
For all $\tau$, the magnitude of the current $|J_{st}|$ first increases sharply with the applied noise. As the noise becomes stronger, $|J_{st}|$ keeps increasing but at a much slower rate. We have checked this for values up to $\sigma \approx 260 \ \lambda_b$. 

The decrease in $|J_{st}|$ with $\tau$ can be understood with the following qualitative argument. Consider the motion of the particle in a single well, before it makes any jumps. If the particle stays in the well up to time $t_{stay} = t_{up} + t_{down}$, its time averaged velocity over that time period is
\begin{equation}
    \dot x \approx  \frac{d_+ + d_-}{t_{stay}}
\end{equation}
where $t_{up}$ and $t_{down}$ are the times the particle spends on the long side and short side respectively. Both times are stochastic quantities.
Since 
\begin{equation}
   d_- \gg d_+ 
\end{equation}
and
\begin{equation}
    t_{up} \gg t_{down}
\end{equation}

\begin{equation}
    J \sim \frac{d_-}{t_{up}}
\end{equation}
where
\begin{equation}
    t_{up} \sim \frac{d_-}{\lambda - \lambda_b}
\end{equation}
This gives
\begin{equation}
    J \sim \lambda - \lambda_b
\end{equation}

$\lambda_b$ is independent of $\tau$. Therefore, the only variable is the value of $\lambda$. It is reasonable to assume that the value of $\lambda(t)$ at any given time is of the order of the standard deviation $\sigma_\lambda(t)$.

\begin{equation}
    \sigma_\lambda(t) = \sqrt{\sigma (1-e^{-2t/\tau})}
\end{equation}

\begin{equation}
    \lambda(t) \sim \sigma_\lambda(t)
\end{equation}

\begin{figure*}
     \centering
     \begin{subfigure}[b]{0.45\textwidth}
         \centering
         \includegraphics[width=\textwidth]{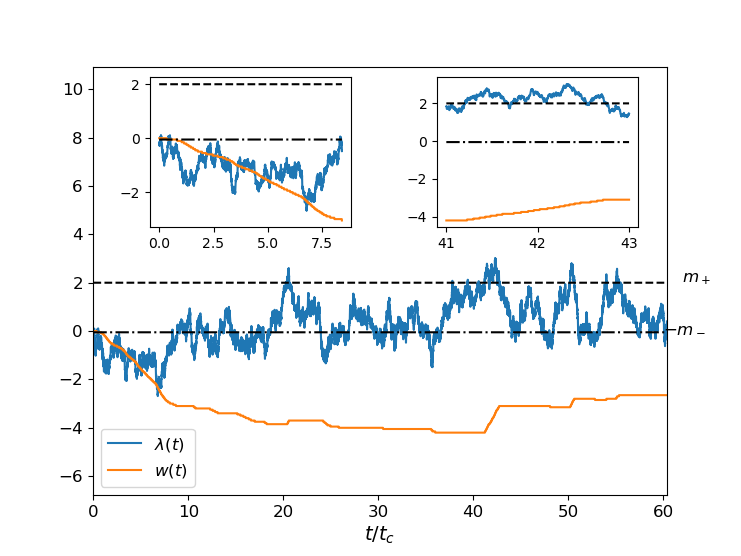}
         \end{subfigure}
     \begin{subfigure}[b]{0.45\textwidth}
         \centering
         \includegraphics[width=\textwidth]{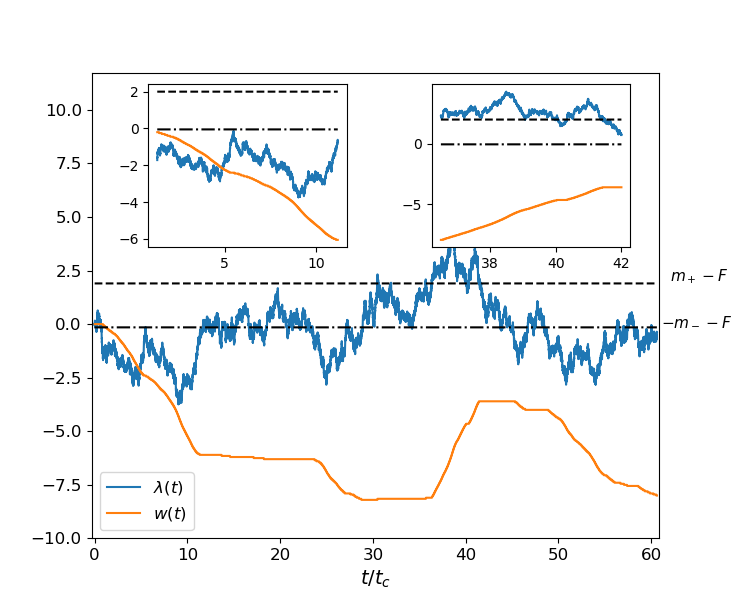}
         \end{subfigure}
     
    \caption{Snapshots showing a typical trajectory of $w(t)$ and $\lambda(t)$ without resetting. (a) without tilting (b) with tilting. The blue curve is $\lambda(t)$ and the orange curve is $w(t)$ which is the instantaneous well number. The inset on the left is an example time interval $T_-$ where $\lambda(t)<\lambda_b$ (dash-dotted line). Here, $w(t)$ decreases with time, corresponding to upward motion. The right inset is an example time interval $T_+$ where $\lambda(t)>\lambda_f$ (dashed line). Well number $w(t)$ increases with time, corresponding to downward motion.}
    \label{fig: one_traj}
    \end{figure*}
    
Note that we cannot use the stationary value of $\sigma_\lambda$ because we keep resetting $\lambda$ at each jump and the distribution of $\lambda$ does not have enough time to relax to the stationary distribution. Since $\sigma$ is kept fixed in all the simulations, the only $\tau$-dependent term is the exponential term. At any given time, $1 - e^{-2t/\tau}$ decreases as $\tau$ increases. Thus $\lambda(t) \sim \sigma(t)$ decreases as $\tau$ increases and the denominator of Eq (11) decreases, leading to an increase in $t_{up}$. Thus from Eq (10), $J(t)$ decreases as $\tau$ increases.

At first glance, the strictly monotonic increase of the current magnitude as a function of noise strength seems counterintuitive. One would expect that as the noise becomes stronger, the likelihood of upward and downward jumps would become roughly equal, leading to a \textit{decrease} in current magnitude. In other words, as the noise is increased, the distributions of $T_-$ and $T_+$ should become roughly equal. As seen from Fig. \ref{fig: current_reset}, however, this is clearly not the case. The reason can be understood by computing the distribution of the time intervals $T_-$ and $T_+$. We plot these quantities for $\sigma \approx 67 \lambda_b$, $\sigma \approx 167 \lambda_b$ and $\sigma \approx 233 \lambda_b$ in Fig. \ref{fig: res_fpt_dist}. Regardless of the strength of the noise, $T_-$ is always more likely to take large values than $T_+$. Because $\lambda$ is reset before it can achieve the steady state, $P(T_-/t_c)$ and $P(T_+/t_c)$ are never equal. Thus, every time $\lambda$ is reset, it is more likely to evolve to values less than $\lambda_b$ at first than to values greater than $\lambda_f$ and hence an upward jump is more likely to occur before a downward jump. In summary, the average number of upward jumps is always greater than that of downward jumps \textit{irrespective of} $\sigma$ leading to the increase of $J_{st}$ observed in Fig. \ref{fig: current_reset}.

\begin{figure*} 
     \centering
     \begin{subfigure}[b]{0.45\textwidth}
         \centering
         \includegraphics[width=\textwidth]{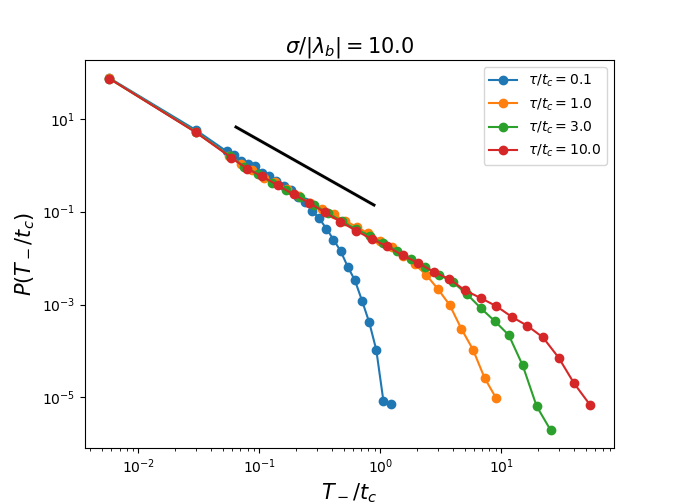}
         \end{subfigure}
     \begin{subfigure}[b]{0.45\textwidth}
         \centering
         \includegraphics[width=\textwidth]{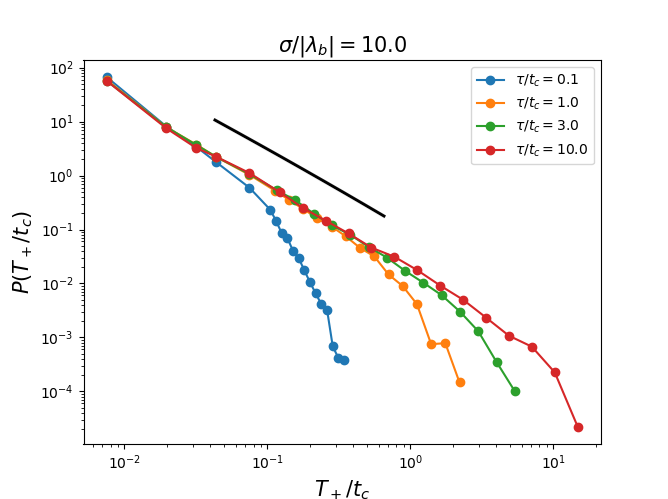}
         \end{subfigure}
         
    \caption{(a) $P(T_-/t_c)$ and (b) $P(T_+/t_c)$ for a range of $\tau$. The tails of both distributions are pushed to larger values with increasing $\tau$.}
    \label{fig: fptd_tau}
    \end{figure*}

\begin{figure*} 
     \centering
     \begin{subfigure}[b]{0.45\textwidth}
         \centering
         \includegraphics[width=\textwidth]{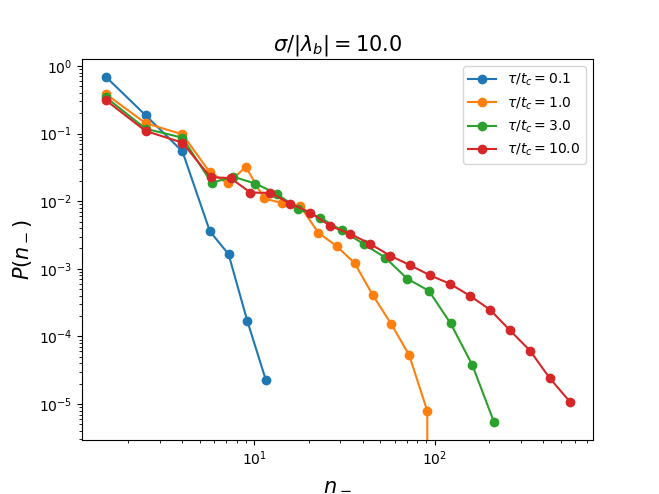}
         \end{subfigure}
     \begin{subfigure}[b]{0.45\textwidth}
         \centering
         \includegraphics[width=\textwidth]{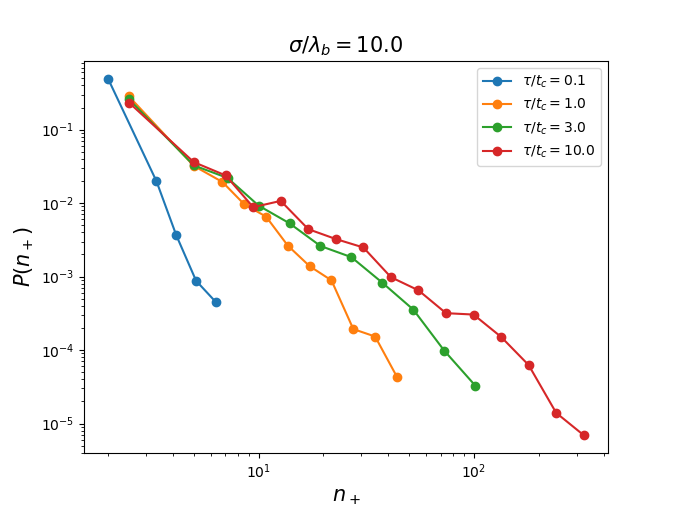}
         \end{subfigure}
         
    \caption{Distributions of (a) $n_-$ (b) $n_+$ for different $\tau$. Just like the distributions of $T_\pm$, the tails of $P(n_\pm)$ are pushed to higher values with increasing $\tau$, with $n_-$ likely to take larger values than $n_+$.}
    \label{fig: n_tau}
    \end{figure*}

\section{particle transport without noise-resetting} \label{section: no-reset}
In the preceding section, we considered a particle hopping between wells in a colored ratchet. At each jump, the value of $\lambda(t)$ was reset to zero. It is interesting to study what happens if $\lambda$ is allowed to evolve according to Eq. (\ref{eq: ou}) without being reset at every jump. As done in the previous section, we show two trajectories, one with $F=0$ and the other with $F=0.1$ (Fig \ref{fig: one_traj}). As before, jumps are seen to occur whenever the threshold values $\lambda_b$ and $\lambda_f$ are crossed. However, in contrast to the previous case, a large number of jumps (upward or downward) can happen successively. Every time $\lambda$ becomes less than (greater than) $\lambda_b \ (\lambda_f)$, a jump occurs shortly after. In the absence of resetting, $\lambda$ stays in below (above) $\lambda_b$ $(\lambda_f)$ for an extended period that depends on $\tau$. In this time interval, many jumps can occur in quick succession, leading to large cascades. 
Another interesting observation is that the larger the time for which $\lambda(t) \geq \lambda_f \ (\lambda(t) \leq \lambda_b)$, the greater is the number of downward (upward) jumps during that time. As before, we denote each interval with $\lambda(t) \geq \lambda_f \ (\lambda(t) \leq \lambda_b)$ by $T_+ \ (T_-)$. $T_+$ and $T_-$ are also random variables. The number of downward (upward) jumps occurring during $T_+ \ (T_-)$ is given by $n_+ \ (n_-)$. The statistics of $T_\pm$ will directly govern the statistics of $n_\pm$. Before calculating the current, it is useful to study the probability densities for $T_+$, $T_-$, $n_-$ and $n_+$.

\subsection{Statistics of $T_-$, $T_+$, $n_-$ and $n_+$}
Mathematically, $T_-$ is the difference between the time at which $\lambda$ crosses just below $\lambda_b$ and the time at which it crosses back above $\lambda_b$ for the \textit{first time}. Similarly, $T_+$ is the difference between the time at which $\lambda$ crosses just above $\lambda_f$ and the time at which it crosses back below $\lambda_f$ for the \textit{first time}. This is precisely the definition of the \textit{first passage time} out of the intervals $(-\infty,\lambda_b]$ and $[\lambda_f, \infty)$. 

The statistics of $T_-$ are obtained by running a single very long simulation of Eqs. (\ref{eq: lang_x}, \ref{eq: ou}) and recording the number and length of time intervals for which $\lambda(t) \leq \lambda_b$. The value of $\lambda$ at time $t=0$ is denoted by $\lambda_0$. The distribution of $T_+$ is obtained in a similar way. The distributions are plotted on a logarithmic scale for several values of $\tau$ in Fig. \ref{fig: fptd_tau}. Both distributions show a power law-like structure with exponent $-3/2$ for intermediate times. There is an exponential cutoff around $T_\pm = \tau$. Hence, the tails of the distributions move to larger values of $T_\pm$ as $\tau$ increases. 

 Second, we observe that $T_-$ is much more likely to take large values than $T_+$ due to the large difference between $\lambda_b$ and $\lambda_f$. In Fig. \ref{fig: n_tau}, we show the distributions of the number of jumps in either direction for varying $\tau$. Again we observe that the tails shift to larger values with increasing $\tau$ in both cases, with $n_-$ generally being larger than $n_+$.

\subsection{Current}

\begin{figure} 
     \centering
     \begin{subfigure}[b]{0.45\textwidth}
         \centering
         \includegraphics[width=\textwidth]{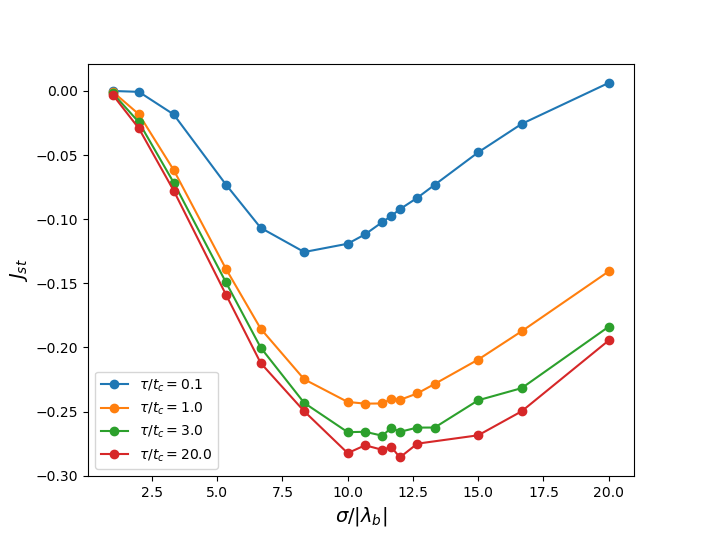}
         \end{subfigure}
     
    \caption{ Stationary particle current $J_{st}$ as a function of $\sigma$ for different $\tau$}
    \label{fig: current}
    \end{figure}

\begin{figure*} 
     \centering
     \begin{subfigure}[b]{0.45\textwidth}
         \centering
         \includegraphics[width=\textwidth]{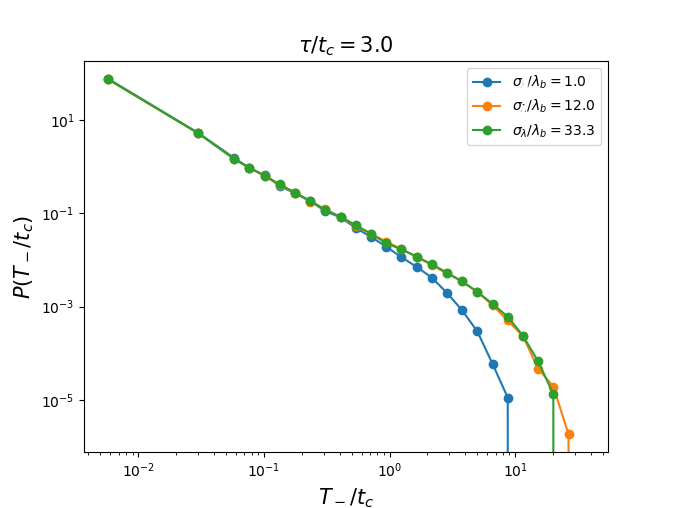}
         \end{subfigure}
     \begin{subfigure}[b]{0.45\textwidth}
         \centering
         \includegraphics[width=\textwidth]{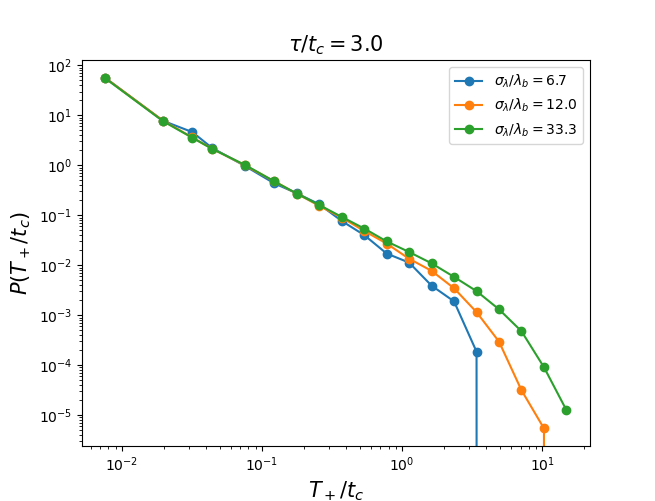}
         \end{subfigure}
         
    \caption{(a) $P(T_-/t_c)$ and (b) $P(T_+/t_c)$ for a range of $\sigma$. }
    \label{fig: fptd_sig}
    \end{figure*}

 We plot the current in Fig \ref{fig: current}. There are two obvious differences with respect to the resetting case. 
 First, for each correlation time there is an optimal noise strength $\sigma_{opt}$ for which the current magnitude is greatest. As seen in Fig. \ref{fig: current}, $\sigma_{opt} \approx 11 \lambda_b$. The second observation is that the current becomes more and more negative as $\tau$ is increased, eventually saturating as $\tau$ increases further. Again, this is in contrast to the resetting case. 

\begin{figure*} 
     \centering
     \begin{subfigure}[b]{0.45\textwidth}
         \centering
         \includegraphics[width=\textwidth]{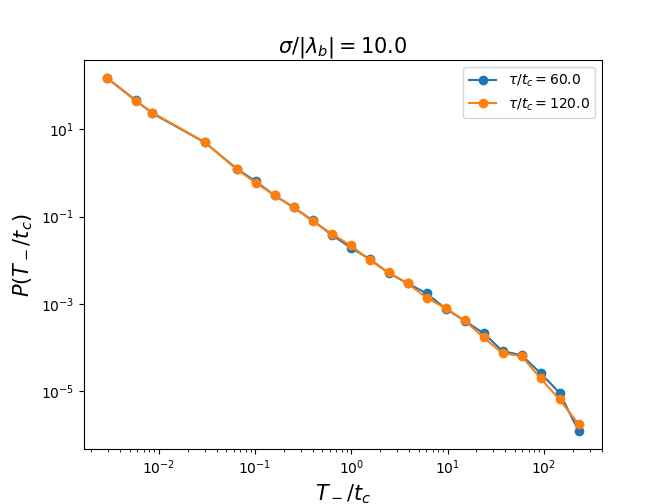}
         \end{subfigure}
     \begin{subfigure}[b]{0.45\textwidth}
         \centering
         \includegraphics[width=\textwidth]{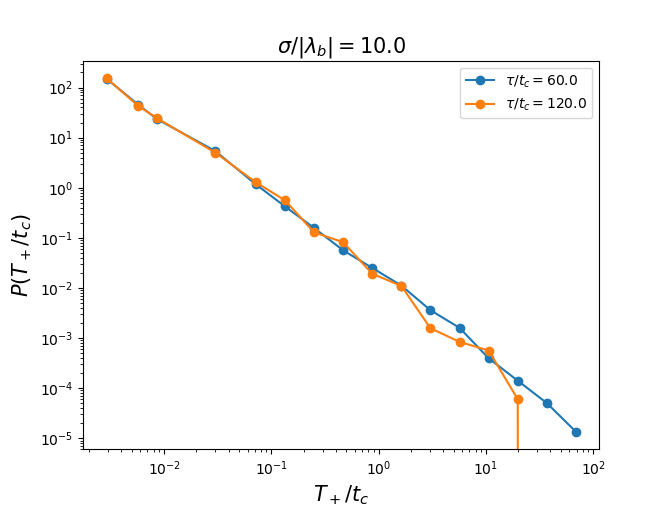}
         \end{subfigure}
         
    \caption{Saturation of the tails of (a) $P(T_-/t_c)$ and (b) $P(T_+/t_c)$ for large $\tau$.}
    \label{fig: fptd_tau_large}
    \end{figure*}

\begin{figure*} 
     \centering
     \begin{subfigure}[b]{0.45\textwidth}
         \centering
         \includegraphics[width=\textwidth]{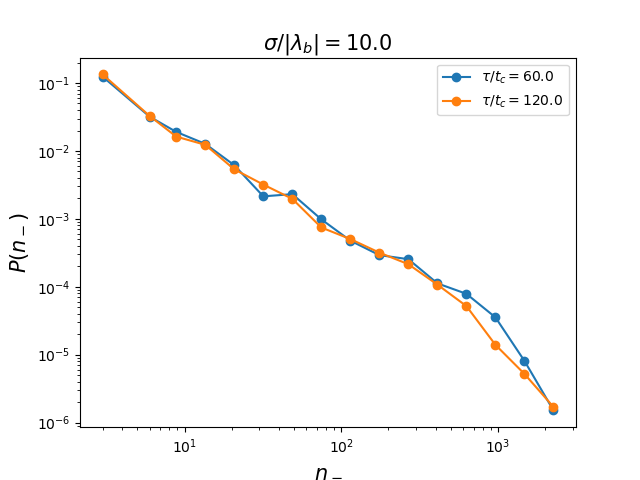}
         \end{subfigure}
     \begin{subfigure}[b]{0.45\textwidth}
         \centering
         \includegraphics[width=\textwidth]{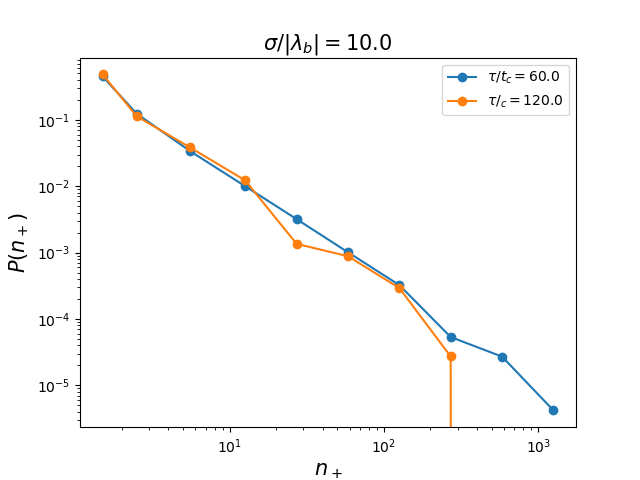}
         \end{subfigure}
         
    \caption{Saturation of the tails of (a) $P(n_-)$ and (b) $P(n_+)$ for large $\tau$. }
    \label{fig: n_tau_large}
    \end{figure*}

\begin{figure} 
     \centering
     \begin{subfigure}[b]{0.45\textwidth}
         \centering
         \includegraphics[width=\textwidth]{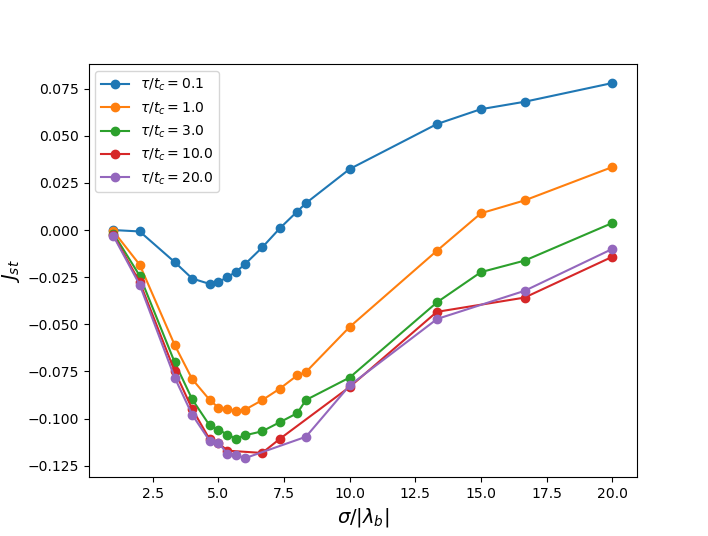}
         \end{subfigure}
    \hfill
     \begin{subfigure}[b]{0.45\textwidth}
         \centering
         \includegraphics[width=\textwidth]{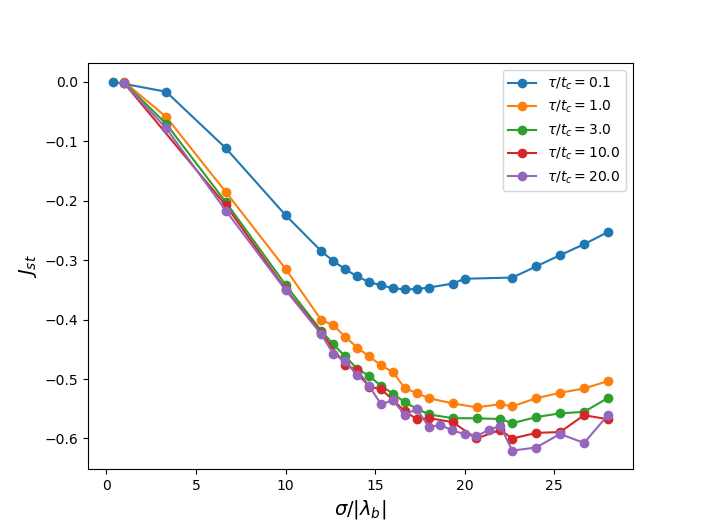}
         \end{subfigure}
     
    \caption{(a) Current versus noise curves for $\lambda_f = 6.5 \ \lambda_b$. (b) Analogous curves for $\lambda_f = 24 \ \lambda_b$. For (a), Eq. (\ref{eq: sig_exp}) gives $\sigma_{exp} \approx 4.7 \ \lambda_b$, and for (b) it gives $\sigma_{exp} \ \approx 13.4 \lambda_b$.} 
    \label{fig: current2}
    \end{figure}
    
\subsection{Optimal noise strength}
To understand the origin of $\sigma_{opt}$, recall that the cascade sizes in both directions depend on how long $\lambda$ remains below $\lambda_b$ or above $\lambda_f$. We study the first passage time distributions $P(T_\pm/t_c)$, this time as a function of $\sigma$. Fig. \ref{fig: fptd_sig} (a) shows $P(T_-/t_c)$ for $\tau = 3 t_c$ and three values of $\sigma$. We see that as $\sigma$ increases beyond $\sigma_{opt}$, the tails of the distributions coincide with each other. In other words, increasing the noise beyond a certain point does not increase the probability of large values of $T_-$. In contrast, Fig. \ref{fig: fptd_sig} (b) shows that the tails of the distributions of $T_+$ move to higher and higher values even after $\sigma$ becomes greater than $\sigma_{opt}$. The net result is that when $\sigma$ exceeds $\sigma_{opt}$, the number of upward jumps $n_-$ stays roughly constant while the number of downward jumps $n_+$ increases. This leads to a decrease in $|J_{st}|$ for $\sigma > \sigma_{opt}$.

We can roughly calculate $\sigma_{opt}$ as follows. It is reasonable to guess that the largest current is obtained when the difference between the the total time below $\lambda_b$ and the total time above $\lambda_f$ is largest. Let $P_b(t)$ be the probability that $\lambda \leq \lambda_b$ at time $t$ and $P_f(t)$ be the probability that $\lambda \geq \lambda_f$ at time $t$. These are found by integrating Eq. (\ref{eq: prob_lbda}) over the appropriate intervals.

\begin{equation}
    P_b(t) = \int_{-\infty}^{\lambda_b} P(\lambda,t) \ d \lambda = 1 + \text{Erf} \left( \frac{\lambda_b - \lambda_0 e^{-t/\tau}}{\sigma (1-e^{-2t/\tau})} \right)
\end{equation}

\begin{equation}
    P_f(t) = \int_{\lambda_f}^\infty P(\lambda, t) \ d \lambda = 1 - \text{Erf} \left( \frac{\lambda_f - \lambda_0 e^{-t/\tau}}{\sigma (1-e^{-2t/\tau})} \right)
\end{equation}

If the simulation is long enough, the fraction of time below and above the thresholds is simply $P_{b, st}$ and $P_{f, st}$ where the subscript $st$ symbolizes the stationary state.

\begin{equation}
    P_{b, st} = \lim_{t \rightarrow \infty} P_b(t) =\frac{1}{2 \sqrt{2}} \left[1 + \text{Erf} \left( \lambda_b/\sigma \right) \right]
\end{equation}
Similarly, 
\begin{equation}
   P_{f, st} = \lim_{t \rightarrow \infty} P_f(t) =\frac{1}{2 \sqrt{2}} \left[1 - \text{Erf} \left( \lambda_f/\sigma \right) \right] 
\end{equation}
Maximizing the quantity $P_{st,b} - P_{st,f}$ with respect to $\sigma$ yields the expected optimum value
\begin{equation} \label{eq: sig_exp}
    \sigma_{exp} = \sqrt{\frac{\lambda_f^2 - \lambda_b^2}{\text{log} \ \lambda_f - \text{log} \ |\lambda_b|} }
\end{equation}
For our parameters, $\sigma_{exp}$ is about $8.2 \lambda_b$. We do not expect this to be very accurate because in reality, jumps do not occur exactly when $\lambda=\lambda_{b}$ or $\lambda=\lambda_f$. There is a finite time lag before the jumps take place. Our reasoning thus underestimates $\sigma_{opt}$. Our formula is also independent of $\tau$, which is only valid for large $\tau$ according to Fig. \ref{fig: current}. Nevertheless, our reasoning explains the existence of an optimal noise and gives us an estimate of its value.

\subsection{Saturation of current curves at large $\tau$}
To understand the effect of $\tau$ on $J_{st}$, we need to understand how the distributions of the two first passage times change as $\tau$ increases.  Unfortunately, a closed expression for $P(T_-/t_c)$ and $P(T_+/t_c)$ is not available, except for the special case where $\lambda_b$ and $\lambda_f$ are zero. Let $T_0$ be the first passage time when $\lambda_{b,f}=0$. Its distribution is given by \cite{ricc}
\begin{multline} \label{eq: fptd}
    F(T_0, \tau) = \frac{|\lambda_0|}{\sqrt{2 D}} \frac{e^{-2T_0/\tau}}{\sqrt{\tau}(1-e^{-2 T_0/\tau})^{3/2}} \ \times \\ \text{exp} \left[ -\frac{\lambda_0^2 e^{-2 T_0/\tau}}{2 D \tau (1-e^{-2 T_0/\tau})}  \right]
\end{multline}
where $\lambda_0$ is the initial value of $\lambda$. This expression has the same qualitative properties as our observed FPTDs and can be used to understand the $\tau \rightarrow \infty$ limit.

\begin{multline}
    \lim_{\tau \rightarrow \infty} F(T_0, \tau) = \lim_{\tau \rightarrow \infty}  \frac{|\lambda_0|}{\sqrt{2 D}} \frac{e^{-2T_0/\tau}}{\sqrt{\tau}(1-e^{-2 T_0/\tau})^{3/2}} \times \\ \lim_{\tau \rightarrow \infty} \text{exp} \left[ -\frac{\lambda_0^2 e^{-2 T_0/\tau}}{2 D \tau (1-e^{-2 T_0/\tau})}  \right]
\end{multline}
Using the fact that $e^{-2 T_0/\tau} \approx 1 - 2 T_0/\tau$ when $\tau \rightarrow \infty$,
the second limit can be shown to be
\begin{equation} \label{eq: lim2}
    \lim_{\tau \rightarrow \infty} \text{exp} \left[ -\frac{\lambda_0^2 e^{-2 T_0/\tau}}{2 D \tau (1-e^{-2T_0/\tau})}  \right] = \text{exp} \left[\frac{-\lambda_0^2}{4 D T_0} \right]
\end{equation}
Therefore, if $T_0$ is kept at some fixed value while $\tau$ increases, the second term approaches the fixed value $e^{-\lambda_0^2/4D T_0}$. On the other hand, for the first term, we can write 
\begin{equation}
   (1-e^{-2T_0/\tau})^{-3/2} \approx{(2 T_0/\tau)^{-3/2}} 
\end{equation}
Substituting this in the denominator and simplifying gives
\begin{multline} \label{eq: lim2}
    \lim_{\tau \rightarrow \infty} \frac{\lambda_0}{\sqrt{2 D}} \frac{e^{-2 T_0/\tau}}{\sqrt{\tau}(1-e^{-2 T_0/\tau})^{3/2}} = \frac{\lambda_0}{\sqrt{2D}} \\ \left( \frac{\tau}{2^{3/2} T_0^{3/2}} - \frac{T_0^{-1/2}}{2^{3/2}} \right)
\end{multline}
Thus the first limit grows linearly with $\tau$, with a slope equal to $\lambda_0/\left( \sqrt{2D} 2^{3/2} T_0^{3/2} \right)$. Since the first limit is constant, the behavior of $F(T_0,\tau)$ is dominated by Eq. (\ref{eq: lim2}). Note that the slope in this equation \textit{decreases} as $T_0$ increases. Therefore, in the tails of $F(T_0, \tau)$, the second limit and hence $F(T_0, \tau)$ are almost independent of $\tau$. In other words, $F(T_0, \tau)$, for large $\tau$ and $T_0$, the FPT distributions saturate (Fig. \ref{fig: fptd_tau_large}) and so do the distributions of $n_\pm$ (Fig. \ref{fig: n_tau_large}). This analysis is not exact for our case, since we have non-zero values of $\lambda_{b,f}$, but it does provide a qualitative reason for why the currents saturate at large $\tau$.

Simulations are repeated for two more values of $m_+$, while keeping $m_-$, $F$ and $d_+ + d_-$ fixed at the original values. Thus $\lambda_b$ stays fixed at 0.15 as in the previous simulations while $\lambda_f$ is changed. In Fig. \ref{fig: current2}, current-noise plots are shown for $\lambda_f = 6.5 \ \lambda_b$ and $\lambda_f=24 \ \lambda_b$. For the first case, the simulations give $\sigma_{opt} \approx 6 \ \lambda_b$, while for the second case, they give $\sigma_{opt} \approx 21 \ \lambda_b$. On the other hand, Eq. (\ref{eq: sig_exp}) yields $\sigma_{exp} \ \approx 4.7 \ \lambda_b$, while for the second case, we get $\sigma_{exp} \approx 13.45 \ \lambda_b$. Eq. (\ref{eq: sig_exp}) thus gives worse results as $\lambda_f$ is increased.

\section{Summary and conclusions} \label{sec: summary}
We have studied the stochastic dynamics of a particle moving in a tilted correlation ratchet for two cases. 
In the first case, the Ornstein-Uhlenbeck process is reset to zero when the particle makes a jump. The current is found to be negative, implying motion against the potential. Interestingly, the current magnitude \textit{decreases} as a function of the correlation time and is a monotonically increasing function of the noise strength. 

In the second case, the net displacement of the particle is again found to be against the tilt. The first passage time distributions of the Ornstein-Uhlenbeck determine the net displacement of the particle and an optimal noise strength is found for which the current is maximum. In contrast to the previous case, the current \textit{increases} with the correlation time of the Ornstein-Uhlenbeck process and saturates for large values. We anticipate that our findings will be useful in understanding how work is done in active matter. The case of cell rearrangements in tissues is an especially promising example. It would be interesting to see if varying the correlation time and noise strength in the vertex model of tissues leads to similar behavior as our toy model.

\bibliography{references}

\begin{thebibliography}{19}%
\makeatletter
\providecommand \@ifxundefined [1]{%
 \@ifx{#1\undefined}
}%
\providecommand \@ifnum [1]{%
 \ifnum #1\expandafter \@firstoftwo
 \else \expandafter \@secondoftwo
 \fi
}%
\providecommand \@ifx [1]{%
 \ifx #1\expandafter \@firstoftwo
 \else \expandafter \@secondoftwo
 \fi
}%
\providecommand \natexlab [1]{#1}%
\providecommand \enquote  [1]{``#1''}%
\providecommand \bibnamefont  [1]{#1}%
\providecommand \bibfnamefont [1]{#1}%
\providecommand \citenamefont [1]{#1}%
\providecommand \href@noop [0]{\@secondoftwo}%
\providecommand \href [0]{\begingroup \@sanitize@url \@href}%
\providecommand \@href[1]{\@@startlink{#1}\@@href}%
\providecommand \@@href[1]{\endgroup#1\@@endlink}%
\providecommand \@sanitize@url [0]{\catcode `\\12\catcode `\$12\catcode `\&12\catcode `\#12\catcode `\^12\catcode `\_12\catcode `\%12\relax}%
\providecommand \@@startlink[1]{}%
\providecommand \@@endlink[0]{}%
\providecommand \url  [0]{\begingroup\@sanitize@url \@url }%
\providecommand \@url [1]{\endgroup\@href {#1}{\urlprefix }}%
\providecommand \urlprefix  [0]{URL }%
\providecommand \Eprint [0]{\href }%
\providecommand \doibase [0]{https://doi.org/}%
\providecommand \selectlanguage [0]{\@gobble}%
\providecommand \bibinfo  [0]{\@secondoftwo}%
\providecommand \bibfield  [0]{\@secondoftwo}%
\providecommand \translation [1]{[#1]}%
\providecommand \BibitemOpen [0]{}%
\providecommand \bibitemStop [0]{}%
\providecommand \bibitemNoStop [0]{.\EOS\space}%
\providecommand \EOS [0]{\spacefactor3000\relax}%
\providecommand \BibitemShut  [1]{\csname bibitem#1\endcsname}%
\let\auto@bib@innerbib\@empty
\bibitem [{\citenamefont {Marchetti}\ \emph {et~al.}(2013)\citenamefont {Marchetti}, \citenamefont {Joanny} \emph {et~al.}}]{marchetti}%
  \BibitemOpen
  \bibfield  {author} {\bibinfo {author} {\bibfnamefont {M.~C.}\ \bibnamefont {Marchetti}}, \bibinfo {author} {\bibfnamefont {J.~F.}\ \bibnamefont {Joanny}}, \emph {et~al.},\ }\href@noop {} {\bibfield  {journal} {\bibinfo  {journal} {Rev. Mod. Phys.}\ }\textbf {\bibinfo {volume} {85}},\ \bibinfo {pages} {1143} (\bibinfo {year} {2013})}\BibitemShut {NoStop}%
\bibitem [{\citenamefont {Chou}\ \emph {et~al.}(2011)\citenamefont {Chou}, \citenamefont {Mallick},\ and\ \citenamefont {Zia}}]{tchou}%
  \BibitemOpen
  \bibfield  {author} {\bibinfo {author} {\bibfnamefont {T.}~\bibnamefont {Chou}}, \bibinfo {author} {\bibfnamefont {K.}~\bibnamefont {Mallick}},\ and\ \bibinfo {author} {\bibfnamefont {R.~K.~P.}\ \bibnamefont {Zia}},\ }\href@noop {} {\bibfield  {journal} {\bibinfo  {journal} {Rep. Prog. Phys.}\ }\textbf {\bibinfo {volume} {74}},\ \bibinfo {pages} {116601} (\bibinfo {year} {2011})}\BibitemShut {NoStop}%
\bibitem [{\citenamefont {Reimann}(2002)}]{reimann}%
  \BibitemOpen
  \bibfield  {author} {\bibinfo {author} {\bibfnamefont {P.}~\bibnamefont {Reimann}},\ }\href@noop {} {\bibfield  {journal} {\bibinfo  {journal} {Phys. Reports}\ }\textbf {\bibinfo {volume} {361}},\ \bibinfo {pages} {57} (\bibinfo {year} {2002})}\BibitemShut {NoStop}%
\bibitem [{\citenamefont {Wang}\ and\ \citenamefont {Nathans}(2007)}]{wang}%
  \BibitemOpen
  \bibfield  {author} {\bibinfo {author} {\bibfnamefont {Y.}~\bibnamefont {Wang}}\ and\ \bibinfo {author} {\bibfnamefont {J.}~\bibnamefont {Nathans}},\ }\href@noop {} {\bibfield  {journal} {\bibinfo  {journal} {Development}\ }\textbf {\bibinfo {volume} {134}},\ \bibinfo {pages} {647} (\bibinfo {year} {2007})}\BibitemShut {NoStop}%
\bibitem [{\citenamefont {Bosveld}\ \emph {et~al.}(2012)\citenamefont {Bosveld}, \citenamefont {Bonnet} \emph {et~al.}}]{bosveld}%
  \BibitemOpen
  \bibfield  {author} {\bibinfo {author} {\bibfnamefont {F.}~\bibnamefont {Bosveld}}, \bibinfo {author} {\bibfnamefont {I.}~\bibnamefont {Bonnet}}, \emph {et~al.},\ }\href@noop {} {\bibfield  {journal} {\bibinfo  {journal} {Science}\ }\textbf {\bibinfo {volume} {336}},\ \bibinfo {pages} {724} (\bibinfo {year} {2012})}\BibitemShut {NoStop}%
\bibitem [{\citenamefont {Duclut}\ \emph {et~al.}(2022)\citenamefont {Duclut}, \citenamefont {Paijmans} \emph {et~al.}}]{duclut}%
  \BibitemOpen
  \bibfield  {author} {\bibinfo {author} {\bibfnamefont {C.}~\bibnamefont {Duclut}}, \bibinfo {author} {\bibfnamefont {J.}~\bibnamefont {Paijmans}}, \emph {et~al.},\ }\href@noop {} {\bibfield  {journal} {\bibinfo  {journal} {Eur. Phys. J. E}\ }\textbf {\bibinfo {volume} {45}},\ \bibinfo {pages} {29} (\bibinfo {year} {2022})}\BibitemShut {NoStop}%
\bibitem [{\citenamefont {Alt}\ \emph {et~al.}(2017)\citenamefont {Alt}, \citenamefont {Ganguly},\ and\ \citenamefont {Salbreux}}]{alt}%
  \BibitemOpen
  \bibfield  {author} {\bibinfo {author} {\bibfnamefont {S.}~\bibnamefont {Alt}}, \bibinfo {author} {\bibfnamefont {P.}~\bibnamefont {Ganguly}},\ and\ \bibinfo {author} {\bibfnamefont {G.}~\bibnamefont {Salbreux}},\ }\href@noop {} {\bibfield  {journal} {\bibinfo  {journal} {Phil. Trans. R. Soc. B}\ }\textbf {\bibinfo {volume} {372}},\ \bibinfo {pages} {20150520} (\bibinfo {year} {2017})}\BibitemShut {NoStop}%
\bibitem [{\citenamefont {Cohen-Addad}\ \emph {et~al.}(2013)\citenamefont {Cohen-Addad}, \citenamefont {Hohler},\ and\ \citenamefont {Pitois}}]{cohenaddad}%
  \BibitemOpen
  \bibfield  {author} {\bibinfo {author} {\bibfnamefont {S.}~\bibnamefont {Cohen-Addad}}, \bibinfo {author} {\bibfnamefont {R.}~\bibnamefont {Hohler}},\ and\ \bibinfo {author} {\bibfnamefont {O.}~\bibnamefont {Pitois}},\ }\href@noop {} {\bibfield  {journal} {\bibinfo  {journal} {Annu. Rev. Fluid. Mech.}\ }\textbf {\bibinfo {volume} {45}},\ \bibinfo {pages} {241} (\bibinfo {year} {2013})}\BibitemShut {NoStop}%
\bibitem [{\citenamefont {Biance}\ \emph {et~al.}(2011)\citenamefont {Biance}, \citenamefont {Delbos},\ and\ \citenamefont {Pitois}}]{biance}%
  \BibitemOpen
  \bibfield  {author} {\bibinfo {author} {\bibfnamefont {A.~L.}\ \bibnamefont {Biance}}, \bibinfo {author} {\bibfnamefont {A.}~\bibnamefont {Delbos}},\ and\ \bibinfo {author} {\bibfnamefont {O.}~\bibnamefont {Pitois}},\ }\href@noop {} {\bibfield  {journal} {\bibinfo  {journal} {Phys. Rev. Lett.}\ }\textbf {\bibinfo {volume} {106}},\ \bibinfo {pages} {068301} (\bibinfo {year} {2011})}\BibitemShut {NoStop}%
\bibitem [{\citenamefont {Keller}\ \emph {et~al.}(2000)\citenamefont {Keller}, \citenamefont {Davidson} \emph {et~al.}}]{keller}%
  \BibitemOpen
  \bibfield  {author} {\bibinfo {author} {\bibfnamefont {R.}~\bibnamefont {Keller}}, \bibinfo {author} {\bibfnamefont {L.}~\bibnamefont {Davidson}}, \emph {et~al.},\ }\href@noop {} {\bibfield  {journal} {\bibinfo  {journal} {Phil. Trans. R. Soc. Lond. B}\ }\textbf {\bibinfo {volume} {355}},\ \bibinfo {pages} {897} (\bibinfo {year} {2000})}\BibitemShut {NoStop}%
\bibitem [{\citenamefont {Tada}\ and\ \citenamefont {Heisenberg}(2012)}]{tada}%
  \BibitemOpen
  \bibfield  {author} {\bibinfo {author} {\bibfnamefont {M.}~\bibnamefont {Tada}}\ and\ \bibinfo {author} {\bibfnamefont {C.-P.}\ \bibnamefont {Heisenberg}},\ }\href@noop {} {\bibfield  {journal} {\bibinfo  {journal} {Development}\ }\textbf {\bibinfo {volume} {139}},\ \bibinfo {pages} {3897} (\bibinfo {year} {2012})}\BibitemShut {NoStop}%
\bibitem [{\citenamefont {Bi}\ \emph {et~al.}(2016)\citenamefont {Bi}, \citenamefont {Yang}, \citenamefont {Marchetti},\ and\ \citenamefont {Manning}}]{bi3}%
  \BibitemOpen
  \bibfield  {author} {\bibinfo {author} {\bibfnamefont {D.}~\bibnamefont {Bi}}, \bibinfo {author} {\bibfnamefont {X.}~\bibnamefont {Yang}}, \bibinfo {author} {\bibfnamefont {M.~C.}\ \bibnamefont {Marchetti}},\ and\ \bibinfo {author} {\bibfnamefont {M.~L.}\ \bibnamefont {Manning}},\ }\href@noop {} {\bibfield  {journal} {\bibinfo  {journal} {Phys. Rev. X}\ }\textbf {\bibinfo {volume} {6}},\ \bibinfo {pages} {021011} (\bibinfo {year} {2016})}\BibitemShut {NoStop}%
\bibitem [{\citenamefont {Fletcher}\ \emph {et~al.}(2014)\citenamefont {Fletcher}, \citenamefont {Osterfield}, \citenamefont {Baker},\ and\ \citenamefont {Shvartsman}}]{fletch}%
  \BibitemOpen
  \bibfield  {author} {\bibinfo {author} {\bibfnamefont {A.~G.}\ \bibnamefont {Fletcher}}, \bibinfo {author} {\bibfnamefont {M.}~\bibnamefont {Osterfield}}, \bibinfo {author} {\bibfnamefont {R.~E.}\ \bibnamefont {Baker}},\ and\ \bibinfo {author} {\bibfnamefont {S.~Y.}\ \bibnamefont {Shvartsman}},\ }\href@noop {} {\bibfield  {journal} {\bibinfo  {journal} {Biophys. J.}\ }\textbf {\bibinfo {volume} {106}},\ \bibinfo {pages} {2291} (\bibinfo {year} {2014})}\BibitemShut {NoStop}%
\bibitem [{\citenamefont {Tetley}\ \emph {et~al.}(2019)\citenamefont {Tetley}, \citenamefont {Staddon}, \citenamefont {Heller}, \citenamefont {Hoppe}, \citenamefont {Banerjee},\ and\ \citenamefont {Mao}}]{tetley}%
  \BibitemOpen
  \bibfield  {author} {\bibinfo {author} {\bibfnamefont {R.~J.}\ \bibnamefont {Tetley}}, \bibinfo {author} {\bibfnamefont {M.~F.}\ \bibnamefont {Staddon}}, \bibinfo {author} {\bibfnamefont {D.}~\bibnamefont {Heller}}, \bibinfo {author} {\bibfnamefont {A.}~\bibnamefont {Hoppe}}, \bibinfo {author} {\bibfnamefont {S.}~\bibnamefont {Banerjee}},\ and\ \bibinfo {author} {\bibfnamefont {Y.}~\bibnamefont {Mao}},\ }\href@noop {} {\bibfield  {journal} {\bibinfo  {journal} {Nat. Phys.}\ }\textbf {\bibinfo {volume} {15}},\ \bibinfo {pages} {1195} (\bibinfo {year} {2019})}\BibitemShut {NoStop}%
\bibitem [{\citenamefont {Commelles}\ \emph {et~al.}(2021)\citenamefont {Commelles}, \citenamefont {Soumya}, \citenamefont {Lu}, \citenamefont {le~Maout}, \citenamefont {Anvitha}, \citenamefont {Salbreux} \emph {et~al.}}]{commelles}%
  \BibitemOpen
  \bibfield  {author} {\bibinfo {author} {\bibfnamefont {J.}~\bibnamefont {Commelles}}, \bibinfo {author} {\bibfnamefont {S.~S.}\ \bibnamefont {Soumya}}, \bibinfo {author} {\bibfnamefont {L.}~\bibnamefont {Lu}}, \bibinfo {author} {\bibfnamefont {E.}~\bibnamefont {le~Maout}}, \bibinfo {author} {\bibfnamefont {S.}~\bibnamefont {Anvitha}}, \bibinfo {author} {\bibfnamefont {G.}~\bibnamefont {Salbreux}}, \emph {et~al.},\ }\href@noop {} {\bibfield  {journal} {\bibinfo  {journal} {eLife}\ }\textbf {\bibinfo {volume} {10}},\ \bibinfo {pages} {e57730} (\bibinfo {year} {2021})}\BibitemShut {NoStop}%
\bibitem [{\citenamefont {Yamamoto}\ \emph {et~al.}(2022)\citenamefont {Yamamoto}, \citenamefont {Sussman}, \citenamefont {Shibata},\ and\ \citenamefont {Manning}}]{yama}%
  \BibitemOpen
  \bibfield  {author} {\bibinfo {author} {\bibfnamefont {T.}~\bibnamefont {Yamamoto}}, \bibinfo {author} {\bibfnamefont {D.~M.}\ \bibnamefont {Sussman}}, \bibinfo {author} {\bibfnamefont {T.}~\bibnamefont {Shibata}},\ and\ \bibinfo {author} {\bibfnamefont {M.~L.}\ \bibnamefont {Manning}},\ }\href@noop {} {\bibfield  {journal} {\bibinfo  {journal} {Soft Matter}\ }\textbf {\bibinfo {volume} {18}},\ \bibinfo {pages} {2168} (\bibinfo {year} {2022})}\BibitemShut {NoStop}%
\bibitem [{\citenamefont {Gardiner}(1985)}]{gardiner}%
  \BibitemOpen
  \bibfield  {author} {\bibinfo {author} {\bibfnamefont {C.}~\bibnamefont {Gardiner}},\ }\href@noop {} {}\ (\bibinfo  {publisher} {Springer-Verlag},\ \bibinfo {address} {Berlin},\ \bibinfo {year} {1985})\BibitemShut {NoStop}%
\bibitem [{\citenamefont {Kloeden}\ and\ \citenamefont {Platen}(1992)}]{kloeden}%
  \BibitemOpen
  \bibfield  {author} {\bibinfo {author} {\bibfnamefont {P.~E.}\ \bibnamefont {Kloeden}}\ and\ \bibinfo {author} {\bibfnamefont {E.}~\bibnamefont {Platen}},\ }\href@noop {} {}\ (\bibinfo  {publisher} {Springer},\ \bibinfo {address} {Berlin},\ \bibinfo {year} {1992})\BibitemShut {NoStop}%
\bibitem [{\citenamefont {Ricciardi}\ and\ \citenamefont {Sato}(1988)}]{ricc}%
  \BibitemOpen
  \bibfield  {author} {\bibinfo {author} {\bibfnamefont {L.~M.}\ \bibnamefont {Ricciardi}}\ and\ \bibinfo {author} {\bibfnamefont {S.}~\bibnamefont {Sato}},\ }\href@noop {} {\bibfield  {journal} {\bibinfo  {journal} {J. Appl. Prob}\ }\textbf {\bibinfo {volume} {25}},\ \bibinfo {pages} {43} (\bibinfo {year} {1988})}\BibitemShut {NoStop}%
\end{thebibliography}%

\end{document}